\RequirePackage{ifpdf}
\ifpdf 
\documentclass[pdftex]{sigma}
\else
\documentclass{sigma}
\fi

\newcommand{\rd}{\partial}

\newcommand{\bst}{\boldsymbol{t}}
\newcommand{\bsx}{\boldsymbol{x}}

\newcommand{\bsD}{\boldsymbol{D}}
\newcommand{\calB}{\mathcal{B}}
\newcommand{\calK}{\mathcal{K}}
\newcommand{\calL}{\mathcal{L}}
\newcommand{\calM}{\mathcal{M}}
\newcommand{\calS}{\mathcal{S}}
\newcommand{\calP}{\mathcal{P}}
\newcommand{\calQ}{\mathcal{Q}}

\numberwithin{equation}{section}

\begin{document}
\allowdisplaybreaks

\renewcommand{\PaperNumber}{057}

\FirstPageHeading

\ShortArticleName{Dispersionless Hirota Equations of Two-Component
BKP Hierarchy}

\ArticleName{Dispersionless Hirota Equations\\ of Two-Component
BKP Hierarchy}

\Author{Kanehisa TAKASAKI} 
\AuthorNameForHeading{K. Takasaki}

\Address{Graduate School of Human and Environmental Studies,
Kyoto University,\\
Yoshida, Sakyo, Kyoto 606-8501, Japan}
\Email{\href{mailto:takasaki@math.h.kyoto-u.ac.jp}{takasaki@math.h.kyoto-u.ac.jp}}

\ArticleDates{Received April 04, 2006, in f\/inal form May 02,
2006; Published online May 31, 2006}

\Abstract{The BKP hierarchy has a two-component analogue (the
2-BKP hierarchy).  Dispersionless limit of this multi-component
hierarchy is considered on the level of the $\tau$-func\-tion.
The so called dispersionless Hirota equations are obtained from
the Hirota equations of the $\tau$-function.  These dispersionless
Hirota equations turn out to be equivalent to a~system of
Hamilton--Jacobi equations. Other relevant equations, in
particular, dispersionless Lax equations, can be derived from
these fundamental equations. For comparison, another approach
based on auxiliary linear equations is also presented.}

\Keywords{BKP hierarchy; Hirota equation; dispersionless limit}

\Classification{35Q58; 37K10; 58F07}

\section{Introduction}

Among many variants of the KP hierarchy \cite{Sato-81,SS-83}, the
BKP hierarchy \cite{DJKM-II,DJKM-IV,DJKM-V,DJKM-VI} enjoys a
distinguished status.  First of all, both the KP and BKP
hierarchies are formulated by a sing\-le $\tau$-function
\cite{DJKM-II}. The $\tau$-function of both hierarchies has a
fermionic representation in terms of a~one-component free fermion
system \cite{DJKM-IV}.  The only dif\/ference lies in the type of
fermions (charged or neutral). In the context of algebraic
geometry, the KP hierarchy characterizes Jacobi varieties, and the
BKP hierarchy corresponds to Prym varieties
\cite{DJKM-V,Shiota-Prym,Krichever-Prym}. As regards the Lax
formalism, both hierarchies are formulated by scalar
pseudo-dif\/ferential operators \cite{DJKM-VI}.  From this point
of view, the BKP hierarchy is a reduction of the KP hierarchy.
This relation carries over to the dispersionless limit
\cite{Takasaki-dBKP}.

The subject of this paper stems from the last topic, namely, the
dispersionless BKP hierarchy. As for the general background on
dispersionless limit of integrable hierarchies, we refer to
Takasaki and Takebe's review \cite{TT-review}.

Recently, Bogdanov and Konopelchenko studied the dispersionless
BKP hierarchy from a new approach \cite{BK-dBKP}.  Their approach
is based on the $\bar{\rd}$-dressing method
\cite{KMAR-dbar,KMA-dbar1,KMA-dbar2,BKMA-geneq}, which had been
successfully applied to other dispersionless integrable
hierarchies including the (genus zero) universal Whitham hierarchy
\cite{Krichever-94}. Employing this method, they obtained an
analogue of the so called ``dispersionless Hirota equations'' for
the BKP hierarchy, which turns out to be slightly dif\/ferent from
those of the KP and Toda hierarchies \cite{TT-review,
CK-95,KKMWWZ-01,Zabrodin-01,BMWRZ-01,Teo-03}.

The aim of this paper is to re-examine this issue along the lines
of approach originally taken for the dispersionless Hirota
equations \cite{TT-review,CK-95}.  Namely, we now start from the
Hirota equations for the $\tau$-function itself rather than its
dispersionless limit.  Moreover, we consider the two-component BKP
(2-BKP) hierarchy that includes two copies of the usual
one-component BKP (1-BKP) hierarchy as a subset.

Unlike the 1-BKP hierarchy, the 2-BKP hierarchy is not very well
known in the literature (see Kac and van de Leur's paper
\cite{KvdL-BDKP} for a comprehensive account of this kind of
multi-component integrable hierarchies). The 2-BKP hierarchy was
f\/irst discovered as a system of Hirota
equations~\cite{DJKM-VII}. A Lax representation was later proposed
in an algebro-geometric framework
\cite{Shiota-Prym,Krichever-Prym}, which revealed that this system
is closely related to two-dimensional f\/inite-gap Schr\"odinger
operators and their isospectral deformations (the Novikov--Veselov
hierarchy) \cite{NV-84}.

The 2-BKP hierarchy is rather special among multi-component
integrable hierarchies, because it has a {\it scalar} Lax
formalism as Shiota \cite{Shiota-Prym} and Krichever
\cite{Krichever-Prym} presented. In this respect, the 2-BKP
hierarchy resembles the Toda hierarchy.  The Toda hierarchy can be
interpreted as the two-component KP (2-KP) hierarchy, though it
has a scalar Lax formalism (in terms of dif\/ference operators)
\cite{UT-84}. The scalar Lax formalism of the 2-BKP hierarchy,
however, has a new feature.  Namely, it has {\it two} spatial
dimensions unlike the one-dimensional setting for the KP, Toda and
1-BKP hierarchies.  This is the origin of all complexities that we
shall encounter.

Our main concern lies in the structure of dispersionless Hirota
equations and some other relevant equations in such a
multi-component setting. We f\/irst derive a set of
``dif\/ferential Fay identities'' from the Hirota equations (more
precisely, their generating functional form). We then apply the
standard prescription of dispersionless limit
\cite{TT-review,CK-95} (which is a kind of ``quasi-classical
limit'') to these equations.  The outcome is a set of equations
that may be called dispersionless Hirota equations.  Whereas part
of them coincide with the equations presented by Bogdanov and
Konopelchenko \cite{BK-dBKP},
 other equations are seemingly new and specif\/ic
to the two-component case.  Moreover, we can correctly derive
other relevant equations of the dispersionless hierarchy such as
Hamilton--Jacobi equations, Lax equations, etc.  We shall see that
two distinct Poisson brackets show up in the formulation of
dispersionless Lax equations. They are manifestation of  presence
of two spatial directions mentioned above.

Let us mention here that Konopelchenko and Moro have studied the
dispersionless limit of the Novikov--Veselov hierarchy by means of
$\bar{\rd}$-dressing \cite{KM-NV}.  Our results should be, in
principle, related to theirs.

This paper is organized as follows. Sections 2 and 3 are devoted
to a brief account of the tau function and equations satisf\/ied
by the tau functions.  In Sections 4 and 5, we derive the
dif\/fe\-ren\-tial Fay identities and the dispersionless Hirota
equations. In Sections 6 and 7, we examine their implications to
the Hamilton--Jacobi equations and the dispersionless Lax
equation. In Section 8, these results are compared with the
conventional approach based on auxiliary linear equations.

\section{Tau functions and bilinear equations}

The 2-BKP hierarchy has two sets of time variables
$t_1,t_3,\ldots,t_{2n+1},\ldots$ and
$\bar{t}_1,\bar{t}_3,\ldots,\bar{t}_{2n+1},\ldots$ with odd
indices\footnote{Throughout this paper, the bar ``$\bar{\;}$'' is
used to create a new symbol, meaning nothing like complex
conjugation.  For instance, $t_n$ and $\bar{t}_n$ are understood
to be independent variables.}. For notational convenience, We
collect these variables into two $\infty$-dimensional vectors
\[
  \bst = (t_1,t_3,\ldots), \qquad
  \bar{\bst} = (\bar{t}_1,\bar{t}_3,\ldots).
\]

The $\tau$-functions $\tau = \tau(\bst,\bar{\bst})$ of this
hierarchy are characterized by the fundamental bilinear
equation~\cite{DJKM-VII,KvdL-BDKP}
\begin{gather}
  \oint \frac{dz}{2\pi iz}e^{\xi(\bst'-\bst,z)}
    \tau\big(\bst' - 2[z^{-1}],\, \bar{\bst}'\big)
    \tau\big(\bst + 2[z^{-1}],\, \bst\big) \nonumber\\
\qquad{}=
  \oint \frac{dz}{2\pi iz}e^{\xi(\bar{\bst}'-\bar{\bst},z)}
    \tau\big(\bst',\, \bar{\bst}' - 2[z^{-1}]\big)
    \tau\big(\bst,\, \bar{\bst} + 2[z^{-1}]\big)
  \label{tau-bilin-eq}
\end{gather}
that holds for arbitrary values of $\bst'$, $\bst$, $\bar{\bst}'$
and $\bar{\bst}$.  Both  sides of this equation are understood to
be contour integrals along the circle $|z| = R$ with
suf\/f\/iciently large radius $R$ and anti-clockwise orientation,
i.e.,
\[
  \oint \frac{dz}{2\pi iz}f(z)
  = \oint_{|z|=R}\frac{dz}{2\pi iz}f(z).
\]
Alternatively, these integrals may be thought of as a purely
algebraic operation
\[
  \oint \frac{dz}{2\pi iz}\sum_{j\in{\mathbb Z}}a_j z^j = a_0
\]
for (formal or convergent) Laurent series. $\xi(\bst,z)$ and
$[z^{-1}]$ denote the BKP version
\[
  \xi(\bst,z) = \sum_{n=0}^\infty t_{2n+1}z^{2n+1}, \qquad
  [z^{-1}] = \left(z^{-1},\frac{z^{-3}}{3},\ldots,
                   \frac{z^{-2n-1}}{2n+1},\ldots\right)
\]
of the usual notation for the KP hierarchy.

We can rewrite this bilinear equation to an inf\/inite number of
Hirota equations as follows. By substituting
\[
  t'_n \to t_n - x_n, \qquad
  t_n  \to t_n + x_n, \qquad
  \bar{t}'_n \to \bar{t}_n - \bar{x}_n, \qquad
  \bar{t}_n  \to \bar{t}_n + \bar{x}_n,
\]
(\ref{tau-bilin-eq}) transforms to
\begin{gather*}
  \oint\frac{dz}{2\pi iz}e^{2\xi(\bsx,z)}
    \tau\big(\bst-\bsx-2[z^{-1}],\,\bar{\bst}-\bar{\bsx}\big)
    \tau\big(\bst+\bsx+2[z^{-1}],\,\bar{\bst}+\bar{\bsx}\big)
  \nonumber\\
\qquad{}=
  \oint\frac{dz}{2\pi iz}e^{2\xi(\bar{\bsx},z)}
    \tau\big(\bst-\bsx,\,\bar{\bst}-\bar{\bsx}-2[z^{-1}]\big)
    \tau\big(\bst+\bsx,\,\bar{\bst}+\bar{\bsx}+2[z^{-1}]\big),
\end{gather*}
where $\bsx = (x_1,x_3,\ldots)$ and $\bar{\bsx} =
(\bar{x}_1,\bar{x}_3,\ldots)$.  If we use the Hirota bilinear
operators $D_{t_{2n+1}}$  and~$D_{\bar{t}_{2n+1}}$ that act on the
product of two functions $f(\bst,\bar{\bst})$,
$g(\bst,\bar{\bst})$ as
\begin{gather*}
  D_{t_{2m+1}}f(\bst,\bar{\bst})\cdot g(\bst,\bar{\bst})
  = \rd_{t_{2n+1}}f(\bst,\bar{\bst})\cdot g(\bst,\bar{\bst})
    - f(\bst,\bar{\bst})\cdot\rd_{t_{2n+1}}g(\bst,\bar{\bst}),
  \nonumber\\
  D_{\bar{t}_{2m+1}}f(\bst,\bar{\bst})\cdot g(\bst,\bar{\bst})
  = \rd_{\bar{t}_{2n+1}}f(\bst,\bar{\bst})\cdot g(\bst,\bar{\bst})
    - f(\bst,\bar{\bst})\cdot\rd_{\bar{t}_{2n+1}}g(\bst,\bar{\bst}),
\end{gather*}
the product of the $\tau$-functions in the integral can be
expressed as
\begin{gather*}
  \tau\big(\bst+\bsx+2[z^{-1}],\,\bar{\bst}+\bar{\bsx}\big)
  \tau\big(\bst-\bsx-2[z^{-1}],\,\bar{\bst}-\bar{\bsx}\big)
 \nonumber \\
\qquad{}= \exp\bigl(\xi(2\tilde{\bsD}_{\bst},z)
        + \bsx\cdot\bsD_{\bst}
        + \bar{\bsx}\cdot\bsD_{\bar{\bst}} \bigr)
      \tau(\bst,\bar{\bst})\cdot\tau(\bst,\bar{\bst}), \\
  \tau\big(\bst+\bsx,\,\bar{\bst}+\bar{\bsx}+2[z^{-1}]\big)
  \tau\big(\bst-\bsx,\,\bar{\bst}-\bar{\bsx}-2[z^{-1}]\big)
 \nonumber \\
\qquad{}= \exp\bigl(\xi(2\tilde{\bsD}_{\bar{\bst}},z)
        + \bst\cdot\bsD_{\bst}
        + \bar{\bst}\cdot\bar{\bsD}_{\bar{\bst}} \bigr)
      \tau(\bst,\bar{\bst})\cdot\tau(\bst,\bar{\bst}),
\end{gather*}
where
\begin{gather*}
  \tilde{\bsD_{\bst}}
  = \left(D_{t_1},\frac{1}{3}D_{t_3},\ldots,
      \frac{1}{2n+1}D_{t_{2n+1}},\ldots\right),
  \\
  \tilde{\bsD}_{\bar{\bst}}
  = \left(D_{\bar{t}_1},\frac{1}{3}D_{\bar{t}_3},\ldots,
      \frac{1}{2n+1}D_{\bar{t}_{2n+1}},\ldots\right),
  \\
  \bsx\cdot\bsD_{\bst}
  = \sum_{n=0}^\infty x_{2n+1}D_{t_{2n+1}},
  \qquad
  \bar{\bsx}\cdot\bsD_{\bar{\bst}}
  = \sum_{n=0}^\infty \bar{x}_{2n+1}D_{\bar{t}_{2n+1}}.
\end{gather*}
Let us introduce the polynomials $q_j(\bst)$, $j = 0,1,2,\ldots$
def\/ined by the generating function
\[
  e^{\xi(\bst,z)} = \sum_{j=0}^\infty q_j(\bst)z^j.
\]
We can thereby rewrite (\ref{tau-bilin-eq}) as
\begin{gather*}
  \sum_{j=0}^\infty q_j(2\bsx)q_j(2\tilde{\bsD}_{\bst})
    \exp(\bsx\cdot\bsD_{\bst} + \bar{\bsx}\cdot\bsD_{\bar{\bst}})
    \tau(\bst,\bar{\bst})\cdot\tau(\bst,\bar{\bst})
 \nonumber \\
\qquad{}=
  \sum_{j=0}^\infty q_j(2\bar{\bsx})q_j(2\tilde{\bsD}_{\bar{\bst}})
    \exp(\bsx\cdot\bsD_{\bst} + \bar{\bsx}\cdot\bsD_{\bar{\bst}})
    \tau(\bst,\bar{\bst})\cdot\tau(\bst,\bar{\bst}).
\end{gather*}
Taylor expansion of this equation with respect to $\bsx$ and
$\bar{\bsx}$ gives an inf\/inite number of Hirota equations.  In
other words, this is a generating functional form of those Hirota
equations.

The 2-BKP hierarchy contains two copies of the 1-BKP hierarchy in
the $\bst$- and $\bar{\bst}$-sectors. These subhierarchies show up
by setting $\bar{\bst}' = \bar{\bst}$ or $\bst' = \bst$. If we set
$\bar{\bst}' = \bar{\bst}$, (\ref{tau-bilin-eq}) reduces to
\begin{gather*}
  \oint \frac{dz}{2\pi iz}e^{\xi(\bst'-\bst,z)}
    \tau\big(\bst' - 2[z^{-1}],\, \bar{\bst}\big)
    \tau\big(\bst + 2[z^{-1}],\, \bar{\bst}\big)
  = \tau(\bst',\bar{\bst})\tau(\bst,\bar{\bst}).
\end{gather*}
Similarly, setting $\bst' = \bst$ yields
\begin{gather*}
  \oint \frac{dz}{2\pi iz}e^{\xi(\bar{\bst}'-\bar{\bst},z)}
    \tau(\bst,\, \bar{\bst}' - 2[z^{-1}])
    \tau(\bst,\, \bar{\bst} + 2[z^{-1}])
  = \tau(\bst,\bar{\bst}')\tau(\bst,\bar{\bst}).
\end{gather*}
These equations are exactly the bilinear equations of the 1-BKP
hierarchy \cite{DJKM-II,DJKM-IV}.

\section{Fermionic representation of tau functions}

The $\tau$-functions of the 1-BKP and 2-BKP hierarchies have a
fermionic representation in terms of neutral fermions
\cite{DJKM-IV,KvdL-BDKP,DJKM-VII}. Let $\varphi_j$, $j \in
{\mathbb Z}$, denote the generators of a Clif\/ford algebra with
the anti-commutation relations
\[
  [\varphi_j,\varphi_k]_{+} = (-1)^j\delta_{j+k,0},
\]
and $|0 \rangle$ and $\langle 0|$ the vacuum states of the Fock
and dual Fock spaces that are annihilated by half of these
generators (``annihilation operators'') as
\[
  \langle 0|\varphi_j = 0 \quad \text{for $j > 0$},
  \qquad
  \varphi_j|0\rangle = 0 \quad \text{for $j < 0$}.
\]
This is the fundamental setting of the neutral fermion system.
Vacuum states of this system are two-dimensional.  Namely, there
are another pair of states $|1 \rangle$ and $\langle 1|$ that are
annihilated by the same set of annihilation operators. These
vacuum states are interchanged by the action of
$\sqrt{2}\varphi_0$ as
\[
  \sqrt{2}\varphi_0 | 0\rangle = |1\rangle, \qquad
  \sqrt{2}\varphi_0 | 1\rangle = |0\rangle, \qquad
  \langle 0| \sqrt{2}\varphi_0 = \langle 1|, \qquad
  \langle 1| \sqrt{2}\varphi_0 = \langle 0|.
\]

Now suppose that $g$ is an invertible element of the Clif\/ford
algebra that generates a ``Bogolioubov transformation'' of the
form
\[
  g\varphi_kg^{-1} = \sum_{j\in{\mathbb Z}}a_{jk}\varphi_j
\]
on the linear space spanned by $\phi_h$'s. A typical (generic)
case is the exponential
\[
  g = \exp\left(\frac{1}{2}\sum_{j,k\ge 0}
      a_{jk}\varphi_j\varphi_k\right)
\]
of a fermionic bilinear form.   Such an operator determines a
$\tau$-function of the 2-BKP hierarchy as the vacuum expectation
value
\[
  \tau(\bst,\bar{\bst})
  = \langle 0 | e^{H(\bst)}ge^{-\bar{H}(\bar{\bst})}
    | 0\rangle,
\]
where $H(\bst)$ and $\bar{H}(\bar{\bst})$ are linear combinations
\[
  H(\bst) = \sum_{n=0}^\infty t_{2n+1}H_{2n+1}, \qquad
  \bar{H}(\bar{\bst}) = \sum_{n=0}^\infty \bar{t}_{2n+1}H_{-2n-1}
\]
of the ``Hamiltonians''
\[
  H_{2n+1}
  = \frac{1}{2}\sum_{j\in{\mathbb Z}}
    (-1)^{j+1}\varphi_j\varphi_{-j-2n-1}.
\]
Note that this resembles the fermionic representation of the
$\tau$-function of the Toda hierarchy
\cite{JM-review,Takebe-TodaTau}.

The fundamental bilinear equation (\ref{tau-bilin-eq}) is a
consequence of the algebraic relation
\[
    \sum_{j\in{\mathbb Z}} (-1)^j \varphi_j g \otimes \varphi_{-j} g
  = \sum_{j\in{\mathbb Z}} (-1)^j g \varphi_j \otimes g \varphi_{-j}
\]
of the foregoing operators and the ``boson-fermion
correspondence''
\begin{gather*}
  \sqrt{2}\langle 1| e^{H(\bst)}\varphi(z) |V\rangle
  = X(\bst,z)\langle 0| e^{H(\bst)} |V\rangle,
  \\
  \sqrt{2}\langle W|\varphi(z^{-1})
    e^{-\bar{H}(\bar{\bst})} |1 \rangle
 = X(\bar{\bst},z)\langle W|
        e^{-\bar{H}(\bar{\bst})} |0 \rangle
\end{gather*}
that holds for arbitrary states $\langle W|$, $| V\rangle$ in the
Fock and dual Fock spaces. $\varphi(z)$ and $X(\bst,z)$ denote the
free fermion f\/ield
\[
  \varphi(z) = \sum_{j\in{\mathbb Z}}\varphi_jz^j
\]
and the corresponding vertex operators
\[
  X(\bst,z)
    = e^{\xi(\bst,z)}e^{-2D(\bst,z)}, \qquad
  D(\bst,z)
    = \sum_{n=0}^\infty \frac{z^{-2n-1}}{2n+1}\rd_{t_{2n+1}}.
\]
The operator $D(\bst,z)$ plays a fundamental role in the
formulation of dispersionless Hirota equations.

\section[Differential Fay identities]{Dif\/ferential Fay identities}

Employing the method developed for the KP hierarchy
\cite{TT-review,CK-95}, we can derive four types of
``dif\/ferential Fay identities''. They are derived from
(\ref{tau-bilin-eq}) by the following procedure.
\begin{itemize}\itemsep=0pt
\item[1)] Dif\/ferentiate (\ref{tau-bilin-eq}) by $t'_1$ and set
\[
  \bst' = \bst + 2[\lambda^{-1}] + 2[\mu^{-1}], \qquad
  \bar{\bst}' = \bar{\bst}.
\]
\item[2)] Dif\/ferentiate (\ref{tau-bilin-eq}) by $\bar{t}'_1$ and
set
\[
  \bst' = \bst, \qquad
  \bar{\bst}' = \bar{\bst} + 2[\lambda^{-1}] + 2[\mu^{-1}].
\]
\item[3)] Dif\/ferentiate (\ref{tau-bilin-eq}) by $t'_1$ and set
\[
  \bst' = \bst + 2[\lambda^{-1}], \qquad
  \bar{\bst}' = \bar{\bst} + 2[\mu^{-1}].
\]
\item[4)] Dif\/ferentiate (\ref{tau-bilin-eq}) by $\bar{t}'_1$,
and set
\[
  \bst' = \bst + 2[\lambda^{-1}], \qquad
  \bar{\bst}' = \bar{\bst} + 2[\mu^{-1}].
\]
\end{itemize}
$\lambda$ and $\mu$ are arbitrary complex constants that sit on
the far side of the integral contour, namely, $|\lambda| > R$ and
$|\mu| > R$.

{\bf 1)}  In this case, we obtain a bilinear equation of the form
\begin{gather*}
  \oint\frac{dz}{2\pi iz}
    \frac{(z+\lambda)(z+\mu)}{(z-\lambda)(z-\mu)}
    \bigl(z\tau\big(\bst+2[\lambda^{-1}]+2[\mu^{-1}]-2[z^{-1}],\,
                \bar{\bst}\big)
 \\
\qquad {}
    + (\rd_{t_1}\tau)\big(\bst+2[\lambda^{-1}]+2[\mu^{-1}]-2[z^{-1}],\,
                 \bar{\bst}\big)
    \bigr)\tau\big(\bst + 2[z^{-1}],\,\bar{\bst}\big) \\
\qquad{}= \oint\frac{dz}{2\pi iz}
    (\rd_{t_1}\tau)\big(\bst+2[\lambda^{-1}]+2[\mu^{-1}],\,
               \bar{\bst} - 2[z^{-1}]\big)
    \tau\big(\bst,\,\bar{\bst} + 2[z^{-1}]\big),
\end{gather*}
where $(\rd_{t_1}\tau)(\bst,\bar{\bst})$ denotes the
$t_1$-derivative of $\tau(\bst,\bar{\bst})$. Also the well known
formula
\[
  \sum_{n=0}^\infty \frac{2z^{2n+1}}{2n+1}
  = \log\frac{1 + z}{1 - z}
\]
has been used to rewrite the exponential factor on the left hand
side as
\[
  e^{\xi(\bst'-\bst,z)}
  = e^{\xi(2[\lambda^{-1}]+2[\mu^{-1}],\,z)}
  = \frac{(z + \lambda)(z + \mu)}{(z - \lambda)(z + \mu)}.
\]
Since the integrands on both  sides of this bilinear equation are
meromorphic functions on the far side $|z| > R$ of the integration
contour, the integrals become the sums of their residues at poles.
Whereas poles of the integrand on the left hand side are located
at $z = \lambda,\mu,\infty$, the integrand on the right hand side
has the only pole at $z = \infty$.  We thus obtain a somewhat
messy bilinear equation for special values of the $\tau$-function
and its $t_1$-derivative.  After some algebra, this bilinear
equation boils down to the following relatively simple form that
resembles the dif\/ferential Fay identity of the KP hierarchy
\cite{TT-review,CK-95}:
\begin{gather}
  \frac{\lambda+\mu}{\lambda-\mu}
  \left(\lambda - \mu - \rd_{t_1}\log
    \frac{\tau(\bst+2[\lambda^{-1}],\,\bar{\bst})}
         {\tau(\bst+2[\mu^{-1}],\,\bar{\bst})} \right)
\nonumber \\
\qquad{}= \left(\lambda + \mu - \rd_{t_1}\log
      \frac{\tau(\bst+2[\lambda^{-1}]+2[\mu^{-1}],\,\bar{\bst})}
           {\tau(\bst,\bar{\bst})} \right)
   \frac{\tau(\bst+2[\lambda^{-1}]+2[\mu^{-1}],\,\bst)
         \tau(\bst,\bar{\bst})}
        {\tau(\bst+2[\lambda^{-1}],\,\bar{\bst})
         \tau(\bst+2[\mu^{-1}],\,\bar{\bst})}.
  \label{dFay-eq1}
\end{gather}
Since these calculations are related to the $\bst$-sector only,
the same equation holds for the tau function of the 1-BKP
hierarchy as well.

{\bf 2)} This case is essentially the same as case 1) except that
the roles of $\bst$ and $\bar{\bst}$ are interchanged. The f\/inal
form of the bilinear equation reads
\begin{gather}
  \frac{\lambda+\mu}{\lambda-\mu}
  \left(\lambda - \mu - \rd_{\bar{t}_1}\log
    \frac{\tau(\bst,\,\bar{\bst}+2[\lambda^{-1}])}
         {\tau(\bst,\,\bar{\bst}+2[\mu^{-1}])} \right)
 \nonumber\\
\qquad{}= \left(\lambda + \mu - \rd_{\bar{t}_1}\log
      \frac{\tau(\bst,\,\bar{\bst}+2[\lambda^{-1}]+2[\mu^{-1}])}
           {\tau(\bst,\bar{\bst})} \right)
   \frac{\tau(\bst,\,\bar{\bst}+2[\lambda^{-1}]+2[\mu^{-1}])
         \tau(\bst,\bar{\bst})}
        {\tau(\bst,\,\bar{\bst}+2[\lambda^{-1}])
         \tau(\bst,\,\bar{\bst}+2[\mu^{-1}])}.
  \label{dFay-eq2}
\end{gather}

{\bf 3)}  In this case, we have a bilinear equation of the form
\begin{gather*}
  \oint\frac{dz}{2\pi iz}
    \frac{z+\lambda}{z-\lambda}
    \bigl(z\tau\big(\bst+2[\lambda^{-1}]-2[z^{-1}],\,
                \bar{\bst}+2[\mu^{-1}]\big)
 \\
\qquad{}
   + (\rd_{t_1}\tau)\big(\bst+2[\lambda^{-1}]-2[z^{-1}],\,
                \bar{\bst}+2[\mu^{-1}]\big)
   \bigr)\tau\big(\bst+2[z^{-1}],\,\bar{\bst}\big) \\
\qquad{}= \oint\frac{dz}{2\pi iz}
    \frac{z+\mu}{z-\mu}
    (\rd_{t_1}\tau)\big(\bst+2[\lambda^{-1}],\,\bar{\bst}+2[\mu^{-1}]\big)
    \tau\big(\bst,\,\bar{\bst}+2[z^{-1}]\big).
\end{gather*}
Again by residue calculus, we obtain the following analogue of the
dif\/ferential Fay identity:
\begin{gather}
  \lambda - \rd_{t_1}\log
     \frac{\tau(\bst+2[\lambda^{-1}],\,\bar{\bst})}
          {\tau(\bst,\,\bar{\bst}+2[\mu^{-1}])}
\nonumber \\
\qquad{}= \left(\lambda - \rd_{t_1}\log
      \frac{\tau(\bst+2[\lambda^{-1}],\,\bar{\bst}+2[\mu^{-1}])}
           {\tau(\bst,\bar{\bst})} \right)
     \frac{\tau(\bst+2[\lambda^{-1}],\,\bar{\bst}+2[\mu^{-1}])
           \tau(\bst,\bar{\bst})}
          {\tau(\bst+2[\lambda^{-1}],\,\bar{\bst})
           \tau(\bst+2[\mu^{-1}],\,\bar{\bst})}.
  \label{dFay-eq3}
\end{gather}
Unlike the previous two equations, this equation has no
counterpart in the 1-BKP hierarchy.

{\bf 4)}  This case is parallel to case 3). The roles of $\bst$
and $\bar{\bst}$ are simply interchanged.  The f\/inal form of the
bilinear equation reads
\begin{gather}
  \mu - \rd_{\bar{t}_1}\log
     \frac{\tau(\bst,\,\bar{\bst}+2[\mu^{-1}])}
          {\tau(\bst+2[\lambda^{-1}],\,\bar{\bst})}
 \nonumber \\
\qquad{}= \left(\mu - \rd_{\bar{t}_1}\log
      \frac{\tau(\bst+2[\lambda^{-1}],\,\bar{\bst}+2[\mu^{-1}])}
           {\tau(\bst,\bar{\bst})} \right)
     \frac{\tau(\bst+2[\lambda^{-1}],\,\bar{\bst}+2[\mu^{-1}])
           \tau(\bst,\bar{\bst})}
          {\tau(\bst+2[\lambda^{-1}],\,\bar{\bst})
           \tau(\bst+2[\mu^{-1}],\,\bar{\bst})}.
  \label{dFay-eq4}
\end{gather}
This equation, too, has no counterpart in the 1-BKP hierarchy.

We have thus derived the four types of equations,
(\ref{dFay-eq1})--(\ref{dFay-eq4}), as an analogue of the
dif\/ferential Fay identity of the KP hierarchy.  The latter is
known to be equivalent to the full KP hierarchy \cite[Appendix
B]{TT-review}.  It seems likely that the same is also true for the
2-BKP hierarchy. Namely, we conjecture that the system of
equations (\ref{dFay-eq1})--(\ref{dFay-eq4}) will be equivalent to
the 2-BKP hierarchy.

The subsequent consideration may be thought of as partial evidence
that supports this conjecture. We shall examine the dispersionless
limit of these equations, which will eventually turn out to be
equivalent to the dispersionless limit of the 2-BKP hierarchy.

\section{Dispersionless Hirota equations}

As in the case of other integrable hierarchies \cite{TT-review},
dispersionless limit is realized as a kind of ``quasi-classical
limit''. To consider this limit in the language of
$\tau$-func\-tion, we allow the $\tau$-func\-tion $\tau =
\tau(\hbar,\bst,\bar{\bst})$ to depend on the ``Planck constant''
$\hbar$ and assume that the rescaled $\tau$-function
\[
  \tau_\hbar(\bst,\bar{\bst})
  = \tau(\hbar,\hbar^{-1}\bst,\hbar^{-1}\bar{\bst})
\]
behaves as
\begin{gather}
  \log\tau_\hbar(\bst,\bar{\bst})
  = \hbar^{-2}F(\bst,\bar{\bst}) + O(\hbar^{-1})
  \label{tau-F}
\end{gather}
in the classical limit $\hbar \to 0$. The goal of the following
consideration is to derive a set of dif\/ferential equations for
the scaling function $F = F(\bst,\bar{\bst})$ from the
dif\/ferential Fay identities (\ref{dFay-eq1})--(\ref{dFay-eq4}).

Upon rescaling the time variables as above, the derivative
operators and the bracket symbols in
(\ref{dFay-eq1})--(\ref{dFay-eq4}) are also rescaled as
\[
  \rd_{t_{2n+1}} \to \hbar\rd_{t_{2n+1}}, \quad
  \rd_{\bar{t}_{2n+1}} \to \hbar\rd_{\bar{t}_{2n+1}}, \quad
  [\lambda^{-1}] \to \hbar[\lambda^{-1}], \quad
  [\mu^{-1}] \to \hbar[\mu^{-1}].
\]
Bearing this in mind, let us examine the limit of these equations
as $\hbar \to 0$.

{\bf 1)} Let us f\/irst consider (\ref{dFay-eq1}). By rescaling,
this equation transforms to
\begin{gather}
  \frac{\lambda+\mu}{\lambda-\mu}
  \left(\lambda - \mu - \hbar\rd_{t_1}\log
    \frac{\tau_\hbar(\bst+2\hbar[\lambda^{-1}],\,\bar{\bst})}
         {\tau_\hbar(\bst+2\hbar[\mu^{-1}],\,\bar{\bst})} \right)
 \nonumber \\
\qquad{}= \left(\lambda + \mu - \hbar\rd_{t_1}\log
      \frac{\tau_\hbar(\bst+2\hbar[\lambda^{-1}]+2\hbar[\mu^{-1}],\,
              \bar{\bst})}
           {\tau_\hbar(\bst,\bar{\bst})} \right)
  \nonumber \\
\phantom{\qquad{}=} {} \times
   \frac{\tau_\hbar(\bst+2\hbar[\lambda^{-1}]+2\hbar[\mu^{-1}],\,
           \bar{\bst})\tau_\hbar(\bst,\bar{\bst})}
        {\tau_\hbar(\bst+2\hbar[\lambda^{-1}],\,\bar{\bst})
         \tau_\hbar(\bst+2\hbar[\mu^{-1}],\,\bar{\bst})}.
  \label{rescaled-dFay-eq1}
\end{gather}
This equation contains three types of ``shifted''
$\tau$-functions.  We can use the already mentioned dif\/ferential
operator
\[
  D(z) = D(\bst,z)
  = \sum_{n=0}^\infty \frac{z^{-2n-1}}{2n+1}\rd_{t_{2n+1}}
\]
to express these shifted $\tau$-functions as
\begin{gather*}
  \tau_\hbar(\bst+2\hbar[\lambda^{-1}],\,\bar{\bst})
  = e^{2\hbar D(\lambda)}\tau_\hbar(\bst,\bar{\bst}), \\
  \tau_\hbar(\bst+2\hbar[\mu^{-1}],\,\bar{\bst})
  = e^{2\hbar D(\mu )}\tau_\hbar(\bst,\bar{\bst}), \\
  \tau_\hbar(\bst+2\hbar[\lambda^{-1}]+2\hbar[\mu^{-1}],\,
         \bar{\bst})
  = e^{2\hbar D(\lambda) + 2\hbar D(\mu)}
      \tau_\hbar(\bst,\bar{\bst}).
\end{gather*}
This enables us to evaluate the logarithm of ``$\tau$-quotients''
in the previous equation as
\begin{gather*}
  \log\frac{\tau_\hbar(\bst+2\hbar[\lambda^{-1}],\,\bar{\bst})}
           {\tau_\hbar(\bst+2\hbar[\mu^{-1}],\,\bar{\bst})}
  = \big(e^{2\hbar D(\lambda)} - e^{2\hbar D(\mu)}\big)
      \log\tau_\hbar(\bst,\bar{\bst})
  = 2\hbar^{-1}(D(\lambda) - D(\mu))F + O(1),
\\
  \log\frac{\tau_\hbar(\bst+2\hbar[\lambda^{-1}]
              +2\hbar[\mu^{-1}],\,\bar{\bst})}
           {\tau_\hbar(\bst,\bar{\bst})}
  = \big(e^{2\hbar D(\lambda) + 2\hbar D(\mu)} - 1\big)
      \log\tau_\hbar(\bst,\bar{\bst})\\
\qquad{}
  = 2\hbar^{-1}(D(\lambda) + D(\mu))F + O(1),
\\
  \log\frac{\tau_\hbar(\bst+2\hbar[\lambda^{-1}]+2\hbar[\mu^{-1}],\,
              \bar{\bst})
            \tau_\hbar(\bst,\bar{\bst})}
           {\tau_\hbar(\bst+2\hbar[\lambda^{-1}],\,\bar{\bst})
            \tau_\hbar(\bst+2\hbar[\mu^{-1}],\,\bar{\bst})}
 \nonumber \\
\qquad{}= \big(e^{2\hbar D(\lambda)+2\hbar D(\mu)} + 1
       - e^{2\hbar D(\lambda)} - e^{2\hbar D(\mu)}\big)
      \log\tau_\hbar(\bst,\bar{\bst}) = 4D(\lambda)D(\mu)F + O(\hbar).
\end{gather*}
Having these results, we can readily f\/ind the limit of each
terms in (\ref{rescaled-dFay-eq1}) as $\hbar \to 0$. This yields
the equation
\begin{gather*}
  \frac{\lambda+\mu}{\lambda-\mu}
  (\lambda - \mu - 2D(\lambda)\rd_{t_1}F
   + 2D(\mu)\rd_{t_1}F)
= (\lambda + \mu - 2D(\lambda)\rd_{t_1}F
       - 2D(\mu)\rd_{t_1}F)
      e^{4D(\lambda)D(\mu)F} \!
\end{gather*}
for the $F$-function.   We can rewrite this equation as
\begin{gather}
  \frac{p(\lambda) - p(\mu)}{p(\lambda) + p(\mu)}
  = \frac{\lambda - \mu}{\lambda + \mu}
    e^{4D(\lambda)D(\mu)F},
  \label{dHirota-eq1}
\end{gather}
where $p(z)$ denotes the generating function
\[
  p(z) = z - 2D(z)\rd_{t_1}F
       = z - \sum_{n=0}^\infty \frac{2z^{-2n-1}}{2n+1}
               \rd_{t_{2n+1}}\rd_{t_1}F
\]
of $\rd_{t_{2n+1}}\rd_{t_1}F$. (\ref{dHirota-eq1}) is exactly the
dispersionless Hirota equation derived by Bogdanov and
Konopelchenko \cite{BK-dBKP} for the 1-BKP hierarchy. We can also
introduce the generating function
\[
  S(z) = \sum_{n=0}^\infty t_{2n+1}z^{2n+1} - D(z)F
\]
of the f\/irst derivatives of $F$ to express $p(z)$ as
\[
  p(z) = \rd_{t_1}S(z)
\]
and rewrite  (\ref{dHirota-eq1}) as
\begin{gather*}
  \frac{\rd_{t_1}S(\lambda) - \rd_{t_1}S(\mu)}
       {\rd_{t_1}S(\lambda) + \rd_{t_1}S(\mu)}
  = \frac{\lambda - \mu}{\lambda + \mu}e^{4D(\lambda)D(\mu)F}.
\end{gather*}

{\bf 2)}  We can start from (\ref{dFay-eq2}) in the rescaled form
\begin{gather*}
  \frac{\lambda+\mu}{\lambda-\mu}
  \left(\lambda - \mu - \hbar\rd_{\bar{t}_1}\log
    \frac{\tau_\hbar(\bst,\,\bar{\bst}+2\hbar[\lambda^{-1}])}
         {\tau_\hbar(\bst,\,\bar{\bst}+2\hbar[\mu^{-1}])} \right)
 \nonumber\\
\qquad{}= \left(\lambda + \mu - \hbar\rd_{\bar{t}_1}\log
      \frac{\tau_\hbar(\bst,\,
              \bar{\bst}+2\hbar[\lambda^{-1}]+2\hbar[\mu^{-1}])}
           {\tau_\hbar(\bst,\bar{\bst})} \right)
 \nonumber\\
\phantom{\qquad{}=}{}\times
   \frac{\tau_\hbar(\bst,\,
           \bar{\bst}+2\hbar[\lambda^{-1}]+2\hbar[\mu^{-1}])
         \tau_\hbar(\bst,\bar{\bst})}
        {\tau_\hbar(\bst,\,\bar{\bst}+2\hbar[\lambda^{-1}])
         \tau_\hbar(\bst,\,\bar{\bst}+2\hbar[\mu^{-1}])}
\end{gather*}
and repeat the same calculations as above. The outcome is the
dispersionless Hirota equation
\begin{gather}
  \frac{\bar{p}(\lambda) - \bar{p}(\mu)}
       {\bar{p}(\lambda) + \bar{p}(\mu)}
  = \frac{\lambda - \mu}{\lambda + \mu}
    e^{4\bar{D}(\lambda)\bar{D}(\mu)F}
  \label{dHirota-eq2}
\end{gather}
of the 1-BKP hierarchy in the $\bar{\bst}$-sector. $\bar{D}(z)$
denotes the dif\/ferential operator
\[
  \bar{D}(z) = D(\bar{\bst},z)
  = \sum_{n=0}^\infty \frac{z^{-2n-1}}{2n+1}\rd_{\bar{t}_{2n+1}}
\]
and $\bar{p}(z)$ the generating function
\[
  \bar{p}(z)
  = z - 2\bar{D}(z)\rd_{\bar{t}_1}F
  = z - \sum_{n=0}^\infty \frac{2z^{-2n-1}}{2n+1}
          \rd_{\bar{t}_{2n+1}}\rd_{\bar{t}_1}F
\]
of $\rd_{\bar{t}_{2n+1}}\rd_{\bar{t}_1}F$. We can also use the
generating function
\[
  \bar{S}(z)
  = \sum_{n=0}^\infty \bar{t}_{2n+1}z^{2n+1} - \bar{D}(z)F
\]
of the f\/irst derivatives of $F$ to express $\bar{p}(z)$ as
\[
  \bar{p}(z) = \rd_{\bar{t}_1}\bar{S}(z)
\]
and to rewrite (\ref{dHirota-eq2}) as
\[
  \frac{\rd_{\bar{t}_1}\bar{S}(\lambda) - \rd_{\bar{t}_1}\bar{S}(\mu)}
       {\rd_{\bar{t}_1}\bar{S}(\lambda) + \rd_{\bar{t}_1}\bar{S}(\mu)}
  = \frac{\lambda - \mu}{\lambda + \mu}
    e^{4\bar{D}(\lambda)\bar{D}(\mu)F}.
\]

{\bf 3)}  Let us now turn to (\ref{dFay-eq3}). In the rescaled
form, it reads
\begin{gather*}
  \lambda - \hbar\rd_{t_1}\log
     \frac{\tau_\hbar(\bst+2\hbar[\lambda^{-1}],\,\bar{\bst})}
          {\tau_\hbar(\bst,\,\bar{\bst}+2\hbar[\mu^{-1}])}
 \nonumber \\
\qquad{} =\left(\lambda - \hbar\rd_{t_1}\log
      \frac{\tau_\hbar(\bst+2\hbar[\lambda^{-1}],\,
              \bar{\bst}+2\hbar[\mu^{-1}])}
           {\tau_\hbar(\bst,\bar{\bst})} \right)
     \frac{\tau_\hbar(\bst+2\hbar[\lambda^{-1}],\,
             \bar{\bst}+2\hbar[\mu^{-1}])
           \tau_\hbar(\bst,\bar{\bst})}
          {\tau_\hbar(\bst+2\hbar[\lambda^{-1}],\,\bar{\bst})
           \tau_\hbar(\bst+2\hbar[\mu^{-1}],\,\bar{\bst})}.
\end{gather*}
The logarithm of $\tau$-quotients in this equation can be
evaluated in much the same way as the previous two cases.  We thus
obtain the equation
\[
  \lambda - 2D(\lambda)\rd_{t_1}F + 2\bar{D}(\mu)\rd_{t_1}F
  = (\lambda - 2D(\lambda)\rd_{t_1}F
       - 2\bar{D}(\mu)\rd_{t_1}F)
      e^{4D(\lambda)\bar{D}(\mu)F}
\]
for the $F$-function.  In terms of the $S$-functions, this
equation reads
\begin{gather}
  \frac{\rd_{t_1}S(\lambda) - \rd_{t_1}\bar{S}(\mu)}
       {\rd_{t_1}S(\lambda) + \rd_{t_1}\bar{S}(\mu)}
  = e^{4D(\lambda)\bar{D}(\mu)F}.
  \label{dHirota-eq3}
\end{gather}
Unlike the previous two equations, this equation has no
counterpart in the 1-BKP hierarchy.

{\bf 4)}  We can repeat the same calculations for (\ref{dFay-eq4})
in the rescaled form
\begin{gather*}
  \mu - \hbar\rd_{\bar{t}_1}\log
     \frac{\tau_\hbar(\bst,\,\bar{\bst}+2\hbar[\mu^{-1}])}
          {\tau_\hbar(\bst+2\hbar[\lambda^{-1}],\,\bar{\bst})}
 \nonumber \\
\qquad{}= \left(\mu - \hbar\rd_{\bar{t}_1}\log
      \frac{\tau_\hbar(\bst+2\hbar[\lambda^{-1}],\,
              \bar{\bst}+2\hbar[\mu^{-1}])}
           {\tau_\hbar(\bst,\bar{\bst})} \right)
     \frac{\tau_\hbar(\bst+2\hbar[\lambda^{-1}],\,
             \bar{\bst}+2\hbar[\mu^{-1}])
           \tau_\hbar(\bst,\bar{\bst})}
          {\tau_\hbar(\bst+2\hbar[\lambda^{-1}],\,\bar{\bst})
           \tau_\hbar(\bst+2\hbar[\mu^{-1}],\,\bar{\bst})}
\end{gather*}
to obtain the equation
\begin{gather}
  \frac{\rd_{\bar{t}_1}\bar{S}(\mu) - \rd_{\bar{t}_1}S(\lambda)}
       {\rd_{\bar{t}_1}\bar{S}(\mu) + \rd_{\bar{t}_1}S(\lambda)}
  = e^{4D(\lambda)\bar{D}(\mu)F}.
  \label{dHirota-eq4}
\end{gather}
This equation, like (\ref{dHirota-eq3}), has no counterpart in the
1-BKP hierarchy.

As a f\/inal remark, let us note that the right hand sides of
(\ref{dHirota-eq3}) and (\ref{dHirota-eq4}) coincide. This implies
that the extra equation
\[
    \frac{\rd_{t_1}S(\lambda) - \rd_{t_1}\bar{S}(\mu)}
         {\rd_{t_1}S(\lambda) + \rd_{t_1}\bar{S}(\mu)}
  = \frac{\rd_{\bar{t}_1}\bar{S}(\mu) - \rd_{\bar{t}_1}S(\lambda)}
         {\rd_{\bar{t}_1}\bar{S}(\mu) + \rd_{\bar{t}_1}S(\lambda)}
\]
or, equivalently,
\[
    \rd_{t_1}S(\lambda)\cdot\rd_{\bar{t}_1}S(\lambda)
  = \rd_{t_1}\bar{S}(\mu)\cdot\rd_{\bar{t}_1}\bar{S}(\mu)
\]
holds.  Since $\lambda$ and $\mu$ take arbitrary values, both
sides of the last equation are actually independent of $\lambda$
and $\mu$.  This value can be determined by letting $\lambda,\mu
\to \infty$:
\[
    \lim_{\lambda\to\infty}
    \rd_{t_1}S(\lambda)\cdot\rd_{\bar{t}_1}S(\lambda)
  = \lim_{\mu\to\infty}
    \rd_{t_1}\bar{S}(\mu)\cdot\rd_{\bar{t}_1}\bar{S}(\mu)
  = - 2\rd_{t_1}\rd_{\bar{t}_1}F.
\]
Thus the extra equation reduces to
\begin{gather}
    \rd_{t_1}S(z)\cdot\rd_{\bar{t}_1}S(z)
  = \rd_{t_1}\bar{S}(z)\cdot\rd_{\bar{t}_1}\bar{S}(z)
  = u,
  \label{dHirota-eq5}
\end{gather}
where
\[
  u = - 2\rd_{t_1}\rd_{\bar{t}_1}F.
\]
As it will turn out later, this equation is the Hamilton--Jacobi
equation of a two-dimensional Schr\"odinger equation that
underlies the Novikov--Veselov hierarchy~\cite{NV-84}.

\section[Hamilton-Jacobi equations]{Hamilton--Jacobi equations}

We show here that the dispersionless Hirota equations
(\ref{dHirota-eq1})--(\ref{dHirota-eq4}) are equivalent to a
system of Hamilton--Jacobi equations for the $S$-functions.  As we
shall show later, these Hamilton--Jacobi equations are Lax
equations of the dispersionless 2-BKP hierarchy in disguise. In
other words, this system, too, is a special case of the universal
Whitham hierarchy \cite{Krichever-94}.

Following Teo's idea \cite{Teo-03}, we borrow the notion of
``Faber polynomials'' from the theory of univalent functions (see
Teo's paper \cite{Teo-03} and references cited therein). This
enables us to do calculations in a clear and systematic way. The
Faber polynomials are def\/ined by a generating function, and
eventually turn out to give the Hamiltonians of the
Hamilton--Jacobi equations.

{\bf 1)}   To derive Hamilton--Jacobi equations from
(\ref{dHirota-eq1}), we f\/irst rewrite it to the logarithmic form
\begin{gather}
  \log\frac{p(\lambda)-p(\mu)}{p(\lambda)+p(\mu)}
  = \log\frac{\lambda-\mu}{\lambda+\mu}
    + \sum_{n,m=0}^\infty
      \frac{4\lambda^{-2n-1}\mu^{-2m-1}}{(2n+1)(2m+1)}
      \rd_{t_{2n+1}}\rd_{t_{2m+1}}F.
  \label{log-dHirota-eq1}
\end{gather}
We now introduce the Faber polynomials $\Phi_n(p)$, $n =
1,2,\ldots$, of $p(z)$.  They are def\/ined by a~generating
function of the form
\begin{gather}
  \log\frac{p(z) - w}{z}
  = - \sum_{n=1}^\infty \frac{z^{-n}}{n}\Phi_n(w).
  \label{Phi-def}
\end{gather}
Although this def\/inition is valid for any Laurent series $p(z)$
of the form $p(z) = z + O(1)$, the present setting is special in
that $p(z)$ is an odd function:
\[
  p(-z) = - p(z).
\]
Because of this, we have the identity
\[
  \log\frac{p(-z) + w}{-z} = \log\frac{p(z) - w}{z},
\]
which implies that $\Phi_n(w)$ is even if $n$ is even and odd if
$n$ is odd, i.e.,
\[
  \Phi_n(-w) = (-1)^n \Phi_n(w).
\]
Therefore we have
\begin{gather}
  \log\frac{p(z) - w}{p(z) + w}
  = - \sum_{n=1}^\infty
        \frac{z^{-n}}{n}
        (\Phi_n(w) - \Phi_n(-w))
 = - \sum_{n=0}^\infty
        \frac{2z^{-2n-1}}{2n+1}\Phi_{2n+1}(w).
  \label{Phi-odd}
\end{gather}
Substituting $z = \lambda$ and $w = p(\mu)$ yields
\[
  \log\frac{p(\lambda) - p(\mu)}{p(\lambda) + p(\mu)}
  = - \sum_{n=0}^\infty
      \frac{2\lambda^{-2n-1}}{2n+1}\Phi_{2n+1}(p(\mu)),
\]
which coincides with the left hand side of
(\ref{log-dHirota-eq1}).  As regards the right hand side of
(\ref{log-dHirota-eq1}), we can rewrite the double sum as
\begin{gather*}
  \sum_{m,n=0}^\infty
  \frac{4\lambda^{-2n-1}\mu^{-2m-1}}{(2n+1)(2m+1)}
  \rd_{t_{2n+1}}\rd_{t_{2m+1}}F
= \sum_{n=0}^\infty
      \frac{2\lambda^{-2n-1}}{2n+1}\left(
        \sum_{m=0}^\infty
        \frac{2\mu^{-2m-1}}{2m+1}
        \rd_{t_{2n+1}}\rd_{t_{2m+1}}F \right)
  \nonumber \\
\qquad{}= \sum_{n=0}^\infty
      \frac{2\lambda^{-2n-1}}{2n+1}\bigl(
        \mu^{2n+1} - \rd_{t_{2n+1}}S(\mu) \bigr)
= - \log\frac{\lambda-\mu}{\lambda+\mu}
      - \sum_{n=0}^\infty
          \frac{2\lambda^{-2n-1}}{2n+1}
          \rd_{t_{2n+1}}S(\mu).
\end{gather*}
Thus (\ref{log-dHirota-eq1}) reduces to the following system of
Hamilton--Jacobi equations:
\begin{gather}
  \rd_{t_{2n+1}}S(z)
  = \Phi_{2n+1}(p(z))
  = \Phi_{2n+1}\bigl(\rd_{t_1}S(z)\bigr),
  \qquad n = 0,1,\ldots.
  \label{HJ-eq1}
\end{gather}

{\bf 2)}  Let us now consider (\ref{dHirota-eq2}).  In this case,
we use the Farber polynomials $\bar{\Phi}_n(\bar{p})$, $n =
1,2,\ldots$ of $\bar{p}(z)$ def\/ined by the generating functional
relation
\[
  \log\frac{\bar{p}(z) - w}{z}
  = - \sum_{n=1}^\infty \frac{z^{-n}}{n}\bar{\Phi}_n(w).
\]
Since $\bar{p}(z)$, like $p(z)$, is an odd function of $z$, we
have the generating function
\[
  \log\frac{\bar{p}(z) - w}{\bar{p}(z) + w}
  = - \sum_{n=0}^\infty
      \frac{2z^{-2n-1}}{2n+1}\bar{\Phi}_{2n+1}(w)
\]
of the Faber polynomials with odd indices. The rest of
calculations are almost the same as in the case of 1).  Thus
(\ref{dHirota-eq2}) reduces to the following Hamilton--Jacobi
equations:
\begin{gather}
  \rd_{\bar{t}_{2n+1}}\bar{S}(z)
  = \bar{\Phi}_{2n+1}(\bar{p}(z))
  = \bar{\Phi}_{2n+1}\bigl(\rd_{\bar{t}_1}\bar{S}(z)\bigr),
  \qquad n = 0,1,\ldots.
  \label{HJ-eq2}
\end{gather}

{\bf 3)}  We can deal with (\ref{dHirota-eq3}) in much the same
way as the case of 1). We start from the logarithmic form
\[
   \log\frac{p(\lambda) - \rd_{t_1}\bar{S}(\mu)}
            {p(\lambda) + \rd_{t_1}\bar{S}(\mu)}
  = \sum_{n,m=0}^\infty
    \frac{4\lambda^{-2n-1}\mu^{-2m-1}}{(2n+1)(2m+1)}
    \rd_{t_{2n+1}}\rd_{\bar{t}_{2m+1}}F
\]
of (\ref{dHirota-eq3}).  By (\ref{Phi-odd}), the left hand side
can be expressed as
\[
  \log\frac{p(\lambda) - \rd_{t_1}\bar{S}(\mu)}
           {p(\lambda) + \rd_{t_1}\bar{S}(\mu)}
  = - \sum_{n=0}^\infty
      \frac{2\lambda^{-2n-1}}{2n+1}
      \Phi_{2n+1}\bigl(\rd_{t_1}\bar{S}(\mu)\bigr).
\]
The right hand side can be written as
\[
  \sum_{n,m=0}^\infty
  \frac{4\lambda^{-2n-1}\mu^{-2m-1}}{(2n+1)(2m+1)}
  \rd_{t_{2n+1}}\rd_{\bar{t}_{2m+1}}F
  = - \sum_{n=0}^\infty
        \frac{2\lambda^{-2n-1}}{2n+1}
        \rd_{t_{2n+1}}\bar{S}(\mu)
\]
Thus (\ref{dHirota-eq3}) reduces to Hamilton--Jacobi equations of
the form
\begin{gather}
  \rd_{t_{2n+1}}\bar{S}(z)
  = \Phi_{2n+1}(\rd_{t_1}\bar{S}(z)),
  \qquad n = 0,1,\ldots.
  \label{HJ-eq3}
\end{gather}

{\bf 4)}  (\ref{dHirota-eq4}) can be treated in the same way as
(\ref{dHirota-eq2}), and boils down to the following
Hamilton--Jacobi equations:
\begin{gather}
  \rd_{\bar{t}_{2n+1}}S(z)
  = \bar{\Phi}_{2n+1}(\rd_{\bar{t}_1}S(z)),
  \qquad n = 0,1,2,\ldots.
  \label{HJ-eq4}
\end{gather}

We have thus derived the Hamilton--Jacobi equations
(\ref{HJ-eq1}), (\ref{HJ-eq2}), (\ref{HJ-eq3}) and (\ref{HJ-eq4})
from the dispersionless Hirota equations (\ref{dHirota-eq1}),
(\ref{dHirota-eq2}), (\ref{dHirota-eq3}) and (\ref{dHirota-eq4}).
Since what we have done is simply to expand generating functions,
this procedure can be reversed. The equivalence of these two
systems are thus proven.

As a f\/inal remark, we note that these four sets of
Hamilton--Jacobi equations can be packed into total dif\/ferential
equations of the form
\begin{gather}
  dS(z) = S'(z)dz
    + \sum_{n=0}^\infty
      \Phi_{2n+1}(\rd_{t_1}S(z))dt_{2n+1}
    + \sum_{n=0}^\infty
      \bar{\Phi}_{2n+1}(\rd_{\bar{t}_1}S(z))d\bar{t}_{2n+1},
  \label{dS(z)}
  \\
  d\bar{S}(z) = \bar{S}'(z)dz
    + \sum_{n=0}^\infty
      \Phi_{2n+1}(\rd_{t_1}\bar{S}(z))dt_{2n+1}
    + \sum_{n=0}^\infty
      \bar{\Phi}_{2n+1}(\rd_{\bar{t}_1}\bar{S}(z))d\bar{t}_{2n+1},
  \label{dSbar(z)}
\end{gather}
where $S'(z)$ and $\bar{S}'(z)$ denote the $z$-derivative of
$S(z)$ and $\bar{S}(z)$,
\[
  S'(z) = \frac{\rd S(z)}{\rd z}, \qquad
  \bar{S}'(z) = \frac{\rd\bar{S}(z)}{\rd z}.
\]

\section{Dispersionless Lax equations}

The Hamilton--Jacobi equations (\ref{HJ-eq1}), (\ref{HJ-eq2}),
(\ref{HJ-eq3}) and (\ref{HJ-eq4}) yield a system of
``dispersionless Lax equations'' (i.e., Lax equations with respect
to Poisson brackets rather than usual commutators) for the inverse
functions $\calL(p)$, $\bar{\calL}(\bar{p})$ of $p(z)$,
$\bar{p}(z)$. Since the derivation is almost the same as the case
of the dispersionless KP and Toda hierarchies \cite{TT-review}, we
here show just an outline of the results, specifying some new
aspects.

\subsection{How to derive dispersionless Lax equations}

Let $z = \calL(p)$ denote the inverse function of $p = p(z)$ in a
neighborhood of $z = \infty$. This is an odd function and has
Laurent expansion of the form
\[
  \calL(p) = p + O(p^{-1})
\]
in a neighborhood of $p = \infty$. We now substitute $z =
\calL(p)$ into (\ref{dS(z)}) and note that (\ref{dHirota-eq5})
implies the relation
\[
  \left.\rd_{\bar{t}_1}S(z)\right|_{z=\calL(p)}
  = \left.\frac{u}{\rd_{t_1}S(z)}\right|_{z=\calL(p)}
  = \frac{u}{p}.
\]
This yields the equation
\begin{gather}
  d\calS(p)
  = \calM(p)d\calL(p)
    + pdt_1 + \sum_{n=1}^\infty \Phi_{2n+1}(p)dt_{2n+1}
    + \frac{u}{p}d\bar{t}_1
    + \sum_{n=1}^\infty
      \bar{\Phi}_{2n+1}\Bigl(\frac{u}{p}\Bigr)d\bar{t}_{2n+1},
  \label{dcalS(p)}
\end{gather}
where
\[
  \calS(p) = S(\calL(p)), \qquad
  \calM(p) = S'(\calL(p)).
\]
Similarly, we can consider the inverse function $z =
\bar{\calL}(\bar{p})$ of $\bar{p} = \bar{p}(z)$, which is an odd
function of $\bar{p}$ and has Laurent expansion of the form
\[
  \bar{\calL}(\bar{p}) = \bar{p} + O(\bar{p}^{-1})
\]
in a neighborhood of $\bar{p} = \infty$. Since (\ref{dHirota-eq5})
implies that
\[
  \left.\rd_{t_1}\bar{S}(z)\right|_{z=\bar{\calL}(\bar{p})}
  = \left.\frac{u}{\rd_{\bar{t}_1}\bar{S}(z)}
    \right|_{z=\bar{\calL}(\bar{p})}
  = \frac{u}{\bar{p}},
\]
(\ref{dSbar(z)}) turns into the equation
\begin{gather}
  d\bar{\calS}(\bar{p})
 = \bar{\calM}(\bar{p})d\bar{\calL}(\bar{p})
    + \frac{u}{\bar{p}}dt_1
    + \sum_{n=1}^\infty
      \Phi_{2n+1}\Bigl(\frac{u}{\bar{p}}\Bigr)dt_{2n+1}
    + \bar{p}d\bar{t}_1
    + \sum_{n=1}^\infty \bar{\Phi}_{2n+1}(\bar{p})d\bar{t}_{2n+1},
  \label{dcalSbar(pbar)}
\end{gather}
where
\[
  \bar{\calS}(\bar{p}) = \bar{S}(\bar{\calL}(\bar{p})),
  \qquad
  \bar{\calM}(\bar{p}) = \bar{S}'(\bar{\calL}(\bar{p})).
\]
$\calL(p)$, $\bar{L}(\bar{p})$ and $\calM(p)$,
$\bar{\calM}(\bar{p})$ are dispersionless analogues of the Lax and
Orlov--Schulman ope\-ra\-tors of the KP and Toda hierarchies
\cite{TT-review}.  Actually, as we show below, the present
situation is more complicated than the case of the  KP and Toda
hierarchies because of
 presence of two ``spatial variables''
$t_1$ and $\bar{t}_1$.

$p$ and $\bar{p}$ in (\ref{dcalS(p)}) and (\ref{dcalSbar(pbar)})
are a priori independent. To derive dispersionless Lax equations,
we have to set the algebraic relation
\[
  p\bar{p} = u
\]
between $p$ and $\bar{p}$.  Note here that the Faber polynomials
in the coef\/f\/icients of (\ref{dcalS(p)}) and
(\ref{dcalSbar(pbar)}) thereby coincide as
\[
  \Phi_{2n+1}(p) = \Phi_{2n+1}(u/\bar{p}), \qquad
  \bar{\Phi}_{2n+1}(u/p) = \bar{\Phi}_{2n+1}(\bar{p}).
\]
 Therefore we have the well-def\/ined 1-form
\[
  \Theta
  = pdt_1
  + \sum_{n=1}^\infty \Phi_{2n+1}(p)dt_{2n+1}
  + \bar{p}d\bar{t}_1
  + \sum_{n=1}^\infty \bar{\Phi}_{2n+1}(\bar{p})d\bar{t}_{2n+1}
\]
and the equation
\[
    d\calS(p) - \calM(p)d\calL(p)
  = \Theta
  = d\bar{\calS}(\bar{p}) - \bar{\calM}(\bar{p})d\bar{\calL}(\bar{p})
\]
of 1-forms, from which we obtain the equation
\[
  d\calL(p) \wedge d\calM(p)
  = d\bar{\calL}(\bar{p})\wedge d\bar{\calM}(\bar{p})
\]
of 2-forms.

Having these equations, we can now resort to the standard
procedure \cite{TT-review} to derive dispersionless Lax equations.
Actually, because of the manifest symmetry between $\bst$, $p$,
$\calL(p)$, $\calM(p)$, $\Phi_{2n+1}(p)$ and $\bar{\bst}$,
$\bar{p}$, $\bar{\calL}(\bar{p})$, $\bar{\calM}(\bar{p})$,
$\bar{\Phi}_{2n+1}(\bar{p})$, we can formulate the dispersionless
Lax equations in two dif\/ferent ways.  A formulation is based on
the Poisson bracket
\[
  \{f,g\}_{p,t_1}
  = \frac{\rd f}{\rd p}\frac{\rd g}{\rd t_1}
  - \frac{\rd f}{\rd t_1}\frac{\rd g}{\rd p}
\]
for the canonical pair $(p,t_1)$. We have the dispersionless Lax
equations
\begin{gather}
  \rd_{t_{2n+1}}\calK(p)
    = \{\Phi_{2n+1}(p),\calK(p)\}_{p,t_1},
  \qquad
  \rd_{\bar{t}_{2n+1}}\calK(p)
   = \{\bar{\Phi}_{2n+1}(u/p),\calK(p)\}_{p,t_1}
  \label{dLax-eq1}
\end{gather}
for $\calK(p) = \calL(p)$, $\calM(p)$, $\bar{\calL}(u/p)$,
$\bar{\calM}(u/p)$ and the canonical commutation relations
\[
  \{\calL(p),\calM(p)\}_{p,t_1} = 1, \qquad
  \{\bar{\calL}(u/p),\bar{\calM}(u/p)\}_{p,t_1} = 1.
\]
Another formulation is based on the Poisson bracket
\[
  \{f,g\}_{\bar{p},\bar{t}_1}
  = \frac{\rd f}{\rd\bar{p}}\frac{\rd g}{\rd\bar{t}_1}
  - \frac{\rd f}{\rd\bar{t}_1}\frac{\rd g}{\rd\bar{p}}.
\]
for the canonical pair $(\bar{p},\bar{t}_1)$. We have the
dispersionless Lax equations
\begin{gather}
  \rd_{t_{2n+1}}\bar{\calK}(\bar{p})
    = \{\Phi_{2n+1}(u/\bar{p}),
          \bar{\calK}(\bar{p})\}_{\bar{p},\bar{t}_1},
  \qquad
  \rd_{\bar{t}_{2n+1}}\bar{\calK}(\bar{p})
    = \{\bar{\Phi}_{2n+1}(\bar{p}),
          \bar{\calK}(\bar{p})\}_{\bar{p},\bar{t}_1}
  \label{dLax-eq2}
\end{gather}
for $\bar{\calK}(\bar{p}) = \bar{\calL}(\bar{p})$,
$\bar{\calM}(\bar{p})$, $\calL(u/\bar{p})$, $\calM(u/\bar{p})$ and
the canonical commutation relations
\[
  \{\bar{\calL}(\bar{p}),
    \bar{\calM}(\bar{p})\}_{\bar{p},\bar{t}_1} = 1, \qquad
  \{\calL(u/\bar{p}),
    \calM(u/\bar{p})\}_{\bar{p},\bar{t}_1} = 1.
\]

\subsection{How to determine Hamiltonians}

For these dispersionless Lax equations to form a closed system,
the Hamiltonians have to be written in terms of the $L$-functions.

We can read of\/f such an expression of the Hamiltonians from the
Hamilton--Jacobi equations themselves.   Let us examine
(\ref{HJ-eq1}).  Substituting $z = \calL(p)$ into this equation
yields
\[
  \Phi_{2n+1}(p)
  = \left.\rd_{t_{2n+1}}S(z)\right|_{z=\calL(p)}
  = \calL(p)^{2n+1} + O\bigl(\calL(p)^{-1}\bigr),
\]
which implies that $\Phi_{2n+1}(p)$ coincides with the
``polynomial part'' of $\calL(p)^{2n+1}$:
\begin{gather}
  \Phi_{2n+1}(p) = \bigl(\calL(p)^{2n+1}\bigr)_{\ge 0}.
  \label{Phi-L}
\end{gather}
In the same way, (\ref{HJ-eq4}) yields
\[
  \bar{\Phi}_{2n+1}(\bar{p})
  = \left.\rd_{\bar{t}_{2n+1}}\bar{S}(z)
    \right|_{z=\bar{\calL}(\bar{p})}
  = \bar{\calL}(\bar{p})^{2n+1}
    + O(\bar{p}^{-1}),
\]
so that $\bar{\Phi}_{2n+1}(\bar{p})$ is given by the polynomial
part of the powers of $\bar{\calL}(\bar{p})$:
\begin{gather}
  \bar{\Phi}_{2n+1}(\bar{p})
  = \bigl(\bar{\calL}(\bar{p})^{2n+1}\bigr)_{\ge 0}.
  \label{Phibar-Lbar}
\end{gather}

Actually, (\ref{Phi-L}) and (\ref{Phibar-Lbar}) can be derived
from the def\/inition of the Faber polynomials as well. As regards
$\Phi_{2n+1}(p)$, the def\/ining relation (\ref{Phi-def})
dif\/ferentiated by $z$ yields the identity
\[
  \frac{p'(z)}{p(z) - w}
  = \sum_{n=1}^\infty z^{-n-1}\Phi_n(w).
\]
Therefore, by Cauchy's integral theorem, we have
\[
  \Phi_n(w)
  = \oint_{C}\frac{dz}{2\pi i}\frac{z^np'(z)}{p(z) - w}
  = \oint_{|p|=R} \frac{dp}{2\pi i}\frac{\calL(p)^n}{p - w},
\]
where $C$ is the image of the circle $|p| = R$ by the map $z =
\calL(p)$, being a simple curve for a suf\/f\/iciently large value
of $R$, and $w$ is understood to sit on the near side, i.e., $|w|
< R$. If we insert the Laurent expansion of $\calL(p)$ in this
expression, we obtain the algebraic expression (\ref{Phi-L}). In
particular, (\ref{Phi-L}) holds for the Faber polynomials with
even indices as well, though we do not use them here. Similarly,
we have an integral formula of $\bar{\Phi}_n(\bar{p})$, which
implies the algebraic expression (\ref{Phibar-Lbar}) and its
counterpart for even indices.

\section{Another approach to dispersionless limit}

In this section, we present another approach that was taken in
Takasaki's previous work on the 1-BKP hierarchy
\cite{Takasaki-dBKP}. This approach is based on quasi-classical
(or ``WKB'') approximation of auxiliary linear equations (see the
review \cite{TT-review} and references cited therein).  As
expected, this approach leads to the same Hamilton--Jacobi
equations as derived from the dif\/ferential Fay identities.

\subsection{Wave function and auxiliary linear equations}

Since we have two vertex operators $X(\bst,z)$ and
$X(\bar{\bst},z)$, we can def\/ine two wave functions:
\[
  \Psi(\bst,\bar{\bst},z)
  = \frac{X(\bst,z)\tau(\bst,\bar{\bst})}
         {\tau(\bst,\bar{\bst})},
  \qquad
  \bar{\Psi}(\bst,\bar{\bst},z)
  = \frac{X(\bar{\bst},z)\tau(\bst,\bar{\bst})}
         {\tau(\bst,\bar{\bst})}.
\]
More explicitly,
\[
  \Psi(\bst,\bar{\bst},z) =
    \frac{\tau(\bst-2[z^{-1}],\,\bar{\bst})}
         {\tau(\bst,\bar{\bst})}
    e^{\xi(\bst,z)},
  \qquad
  \bar{\Psi}(\bst,\bar{\bst},z) =
    \frac{\tau(\bst,\,\bar{\bst}-2[z^{-1}])}
         {\tau(\bst,\bar{\bst})}
    e^{\xi(\bar{\bst},z)}.
\]
The bilinear equation (\ref{tau-bilin-eq}) for the tau function
thereby turns into the bilinear equation
\begin{gather}
    \oint\frac{dz}{2\pi iz}
    \Psi(\bst',\bar{\bst}',z)\Psi(\bst,\bar{\bst},-z)
  = \oint\frac{dz}{2\pi iz}
    \bar{\Psi}(\bst',\bar{\bst}',z)\bar{\Psi}(\bst,\bar{\bst},-z)
  \label{wf-bilin-eq}
\end{gather}
for the wave functions.  Note that the role of ``dual wave
functions'' in the KP and Toda hierarchies are now played by
$\Psi(\bst,\bar{\bst},-z)$ and $\bar{\Psi}(\bst,\bar{\bst},-z)$.
Let us mention that such a pair of wave functions for the 2-BKP
hierarchy were f\/irst considered by Shiota \cite{Shiota-Prym} in
an algebro-geometric framework.

As we shall show below, these two wave functions satisfy the
following three sets of ``auxiliary linear equations.''
\begin{itemize}\itemsep=0pt
\item[1.] Linear equations of the 1-BKP type in the $\bst$-sector:
\begin{gather}
  \bigl(\rd_{t_{2n+1}} - B_{2n+1}(\rd_{t_1})\bigr)\Psi = 0,
  \qquad n = 0,1,2,\ldots,
  \label{t-lin-eq}
\end{gather}
where $B_{2n+1}(\partial_{t_1})$ is a dif\/ferential operator in $t_1$ of
the form $\rd_{t_1}^{2n+1} + \cdots$ without $0$-th order term,
i.e.,
\[
  B_{2n+1}(\rd_{t_1})1 = 0.
\]
\item[2.] Linear equations of the 1-BKP type in the
$\bar{\bst}$-sector:
\begin{gather}
  \bigl(\rd_{\bar{t}_{2n+1}} - \bar{B}_{2n+1}(\rd_{\bar{t}_1})
    \bigr)\Psi = 0,
  \qquad n = 0,1,2,\ldots,
  \label{tbar-lin-eq}
\end{gather}
where $\bar{B}_{2n+1}(\rd_{\bar{t}_1})$ is a dif\/ferential
operator in $t_1$ of the form $\rd_{\bar{t}_1}^{2n+1} + \cdots$
without $0$-th order term, i.e.,
\[
  \bar{B}_{2n+1}(\rd_{\bar{t}_1})1 = 0.
\]
\item[3.] Two-dimensional Schr\"odinger equation of the
Novikov--Veselov type \cite{NV-84}:
\begin{gather}
  (\rd_{t_1}\rd_{\bar{t}_1} - u) = 0
  \label{ttbar-lin-eq}
\end{gather}
where $
  u = - 2\rd_{t_1}\rd_{\bar{t}_1}\log\tau.
$
\end{itemize}

Note that the condition on the $0$-th order term of
$B_{2n+1}(\rd_{t_1})$ and $\bar{B}_{2n+1}(\rd_{\bar{t}_1})$ is the
same as the condition that characterizes the $B$-operators of the
1-BKP hierarchy \cite{DJKM-IV,DJKM-VI}.   This is another way to
see that the 2-BKP hierarchy contains two copies of the 1-BKP
hierarchy.

\subsection{How to derive auxiliary linear equations}

These auxiliary linear equations are derived from the bilinear
equation (\ref{wf-bilin-eq}) by the standard method
\cite{DJKM-VI,KvdL-BDKP,DJKM-review,Dickey-book} based on the
following

\begin{lemma}
If $R(\bst,\bar{\bst},z)$ and $\bar{R}(\bst,\bar{\bst},z)$ are a
pair of Laurent series of $z$ of the form
\[
  R(\bst,\bar{\bst},z) = O(z^{-1})e^{\xi(\bst,z)}, \qquad
  \bar{R}(\bst,\bar{\bst},z) = O(z^{-1})e^{\xi(\bar{\bst},z)},
\]
and satisfy the bilinear equation
\[
    \oint \frac{dz}{2\pi iz}
    R(\bst',\bar{\bst}',z)\Psi(\bst,\bar{\bst},-z)
  = \oint \frac{dz}{2\pi iz}
    \bar{R}(\bst',\bar{\bst}',z)\bar{\Psi}(\bst,\bar{\bst},-z),
\]
then $R(\bst,\bar{\bst},z) = \bar{R}(\bst,\bar{\bst},z) = 0$.
\end{lemma}

We refer the proof of this lemma to the references
\cite{DJKM-VI,KvdL-BDKP,DJKM-review,Dickey-book} and illustrate
usage of this lemma for the case of (\ref{t-lin-eq}).
Dif\/ferentiating (\ref{wf-bilin-eq}) by $t_{2n+1}$ yields
\[
  \oint\frac{dz}{2\pi iz}
    \rd_{t'_{2n+1}}\Psi(\bst',\bar{\bst}',z)
    \Psi(\bst,\bar{\bst},-z)
= \oint\frac{dz}{2\pi iz}
      \rd_{t'_{2n+1}}\bar{\Psi}(\bst',\bar{\bst}',z)
      \bar{\Psi}(\bst,\bar{\bst},-z).
\]
Similar bilinear equations hold for higher derivatives as well. In
particular, we have
\[
  \oint\frac{dz}{2\pi iz}
    B(\rd_{t_1})\Psi(\bst',\bar{\bst}',z)
    \Psi(\bst,\bar{\bst},-z)
= \oint\frac{dz}{2\pi iz}
      B(\rd_{t_1})\bar{\Psi}(\bst',\bar{\bst}',z)
      \bar{\Psi}(\bst,\bar{\bst},-z).
\]
for any choice of dif\/ferential operator $B(\rd_{t_1})$ in $t_1$.
This implies that the bilinear equation in the lemma is
satisf\/ied by
\begin{gather*}
  R(\bst,\bar{\bst},z)
  = \rd_{t_{2n+1}}\Psi(\bst,\bar{\bst},z)
    - B(\rd_{t_1})\Psi(\bst,\bar{\bst},z),
  \\
  \bar{R}(\bst,\bar{\bst},z)
 = \rd_{t_{2n+1}}\bar{\Psi}(\bst,\bar{\bst},z)
    - B(\rd_{t_1})\bar{\Psi}(\bst,\bar{\bst},z).
\end{gather*}
If $B(\rd_{t_1})$ is chosen to be an operator
$B_{2n+1}(\rd_{t_1})$ for which other conditions of the lemma are
satisf\/ied, then $R(\bst,\bar{\bst},z)$ and
$\bar{R}(\bst,\bar{\bst},z)$ automatically vanish, thus we obtain
(\ref{t-lin-eq}). The extra condition on the $0$-th order term of
$B_{2n+1}(\rd_{t_1})$ turns out to be satisf\/ied because
(\ref{wf-bilin-eq}) reduces to the bilinear equation of the 1-BKP
hierarchy in the $\bst$-sector upon setting $\bar{\bst}' =
\bar{\bst}$.

(\ref{tbar-lin-eq}) and (\ref{ttbar-lin-eq}) can be derived in the
same way.

\subsection{Lax and zero-curvature equations}

We now consider a ``scalar'' Lax formalism of the 2-BKP hierarchy
on the basis of the auxiliary linear equations presented above.
This Lax formalism is conceptually parallel to the
algebro-geometric framework based on ``two-point'' (or
``two-puncture'') Baker-Akhiezer functions
\cite{Shiota-Prym,Krichever-Prym}.  In spite of some unusual
aspects as we see  below, such a scalar Lax formalism is more
suited to the consideration of dispersionless limit than the
``matrix'' Lax formalism of Kac and van de Leur \cite{KvdL-BDKP}.

Let us f\/irst consider zero-curvature equations for the
$B$-operators.  As pointed out by Kri\-che\-ver
\cite{Krichever-Prym}, the $B$-operators satisfy the modif\/ied
``zero-curvature'' equations
\[
  \bigl[\rd_{t_{2m+1}} - B_{2m+1}(\rd_{t_1}),\,
  \rd_{\bar{t}_{2n+1}} - \bar{B}_{2n+1}(\rd_{\bar{t}_1})\bigr]
  = D_{mn}(\rd_t,\rd_{\bar{t}_1})(\rd_{t_1}\rd_{\bar{t}_1} - u),
\]
where $D_{mn}(\rd_{t_1},\rd_{\bar{t}_1})$ are dif\/ferential
operators in both $t_1$ and $\bar{t}_1$, alongside the usual
zero-curvature equations
\begin{gather*}
  \bigl[\rd_{t_{2m+1}} - B_{2m+1}(\rd_{t_1}),\,
  \rd_{t_{2n+1}} - B_{2n+1}(\rd_{t_1})\bigr]
    = 0, \\
  \bigl[\rd_{\bar{t}_{2m+1}} - \bar{B}_{2m+1}(\rd_{\bar{t}_1}),\,
  \rd_{\bar{t}_{2n+1}} - \bar{B}_{2n+1}(\rd_{\bar{t}_1})\bigr]
    = 0
\end{gather*}
of the 1-BKP hierarchy in the $\bst$- and $\bar{\bst}$-sectors. In
other words, $\rd_{t_{2m+1}} - B_{2m+1}(\rd_{t_1})$ and
$\rd_{\bar{t}_{2n+1}} - \bar{B}_{2n+1}(\rd_{\bar{t}_1})$ commute
``modulo'' the two-dimensional Schr\"odinger operator
$\rd_{t_1}\rd_{\bar{t}_1} - u$.  We refer the details to
Krichever's paper \cite{Krichever-Prym}.

Let us turn to Lax equations. We have two $L$-operators
\[
  L(\rd_{t_1})
    = W(\rd_{t_1})\rd_{t_1}W(\rd_{t_1})^{-1}, \qquad
  \bar{L}(\rd_{\bar{t}_1})
    = \bar{W}(\rd_{\bar{t}_1})\rd_{\bar{t}_1}
      \bar{W}(\rd_{\bar{t}_1})^{-1},
\]
where $W(\rd_{t_1})$ and $\bar{W}(\rd_{\bar{t}_1})$ are dressing
operators of the form
\[
  W(\rd_{t_1})
  = 1 + \sum_{j=1}^\infty
        w_j(\bst,\bar{\bst})\rd_{t_1}^{-j},
  \qquad
  \bar{W}(\rd_{\bar{t}_1})
  = 1 + \sum_{j=1}^\infty
        \bar{w}_j(\bst,\bar{\bst})\rd_{\bar{t}_1}^{-j}
\]
that are related to the wave functions as
\[
  \Psi(\bst,\bar{\bst},z)
    = W(\rd_{t_1})e^{\xi(\bst,z)}, \qquad
  \bar{\Psi}(\bst,\bar{\bst},z)
    = \bar{W}(\rd_{\bar{t}_1})e^{\xi(\bar{\bst},z)}.
\]
Substituting this expression of the wave functions into
(\ref{t-lin-eq}) and (\ref{tbar-lin-eq}) yields the so called Sato
equations
\begin{gather*}
  \frac{\rd W(\rd_{t_1})}{\rd t_{2n+1}}
  = B_{2n+1}(\rd_{t_1})W(\rd_{t_1})
      - W(\rd_{t_1})\rd_{t_1}^{2n+1},  \\
  \frac{\rd\bar{W}(\rd_{\bar{t}_1})}{\rd\bar{t}_{2n+1}}
  = \bar{B}_{2n+1}(\rd_{\bar{t}_1})\bar{W}(\rd_{\bar{t}_1})
      - \bar{W}(\rd_{\bar{t}_1})\rd_{\bar{t}_1}^{2n+1},
\end{gather*}
which in turn imply the following expression of the $B$-operators:
\begin{gather}
  B_{2n+1}(\rd_{t_1})
    = \bigl(L(\rd_{t_1})^{2n+1}\bigr)_{\ge 0},
  \qquad
  \bar{B}_{2n+1}(\rd_{\bar{t}_1})
    = \bigl(\bar{L}(\rd_{\bar{t}_1})^{2n+1}\bigr)_{\ge 0},
  \label{BBbar-LLbar}
\end{gather}
where $(\quad)_{\ge 0}$ denotes the dif\/ferential operator part
of a pseudo-dif\/ferential operator. Since
$\Psi(\bst,\bar{\bst},z)$ and $\bar{\Psi}(\bst,\bar{\bst},z)$
satisfy the linear equations
\begin{gather}
  L(\rd_{t_1})\Psi(\bst,\bar{\bst},z)
    = z\Psi(\bst,\bar{\bst},z), \qquad
  \bar{L}(\bst,\bar{\bst},z)\bar{\Psi}(\bst,\bar{\bst},z)
    = z\bar{\Psi}(\bst,\bar{\bst},z)
  \label{L-lin-eq}
\end{gather}
alongside (\ref{t-lin-eq}) and (\ref{tbar-lin-eq}), we have the
Lax equations
\begin{gather}
  \frac{\rd L(\rd_{t_1})}{\rd t_{2n+1}}
    = [B_{2n+1}(\rd_{t_1}),L(\rd_{t_1})], \qquad
  \frac{\rd\bar{L}(\rd_{\bar{t}_1})}{\rd\bar{t}_{2n+1}}
    = [\bar{B}_{2n+1}(\rd_{\bar{t}_1}),\bar{L}(\rd_{\bar{t}_1})],
  \label{Lax-eq1}
\end{gather}
which give the standard Lax representation of the 1-BKP hierarchy
in the $\bst$- and $\bar{\bst}$-sector, respectively.

It is, however, not so straightforward to derive Lax equations of
the form
\[
 \frac{\rd L(\rd_{t_1})}{\rd\bar{t}_{2n+1}}
   = [*,L(\rd_{t_1})], \qquad
 \frac{\rd\bar{L}(\rd_{\bar{t}_1}) }{\rd t_{2n+1}}
   = [*,\bar{L}(\rd_{\bar{t}_1})],
\]
Those Lax equations can be obtained from auxiliary linear
equations of the form
\begin{gather}
  \bigl(\bar{Q}_{2n+1}(\rd_{t_1})\rd_{\bar{t}_{2n+1}}
      - \bar{P}_{2n+1}(\rd_{t_1}) \bigr)\Psi = 0,
  \qquad
  \bigl(Q_{2n+1}(\rd_{\bar{t}_1})\rd_{t_{2n+1}}
      - P_{2n+1}(\rd_{\bar{t}_1}) \bigr)\Psi = 0,
  \label{ttbar-lin-eq2}
\end{gather}
where $\bar{P}_{2n+1}(\rd_{t_1}),\bar{Q}_{2n+1}(\rd_{t_1})$ and
$P_{2n+1}(\rd_{\bar{t}_1}),Q_{2n+1}(\bar{t}_1)$ are dif\/ferential
operators in $t_1$ and $\bar{t}_1$, respectively.  One can derive
these linear equations by the same reasoning as in the case of
(\ref{ttbar-lin-eq}).  Note that (\ref{ttbar-lin-eq}) amounts to
the case of $n = 1$ with
\[
  \bar{Q}_1(\rd_{t_1}) = \rd_{t_1}, \qquad
  \bar{P}_1(\rd_{t_1}) = u, \qquad
  Q_1(\rd_{\bar{t}_1}) = \rd_{\bar{t}_1}, \qquad
  P_1(\rd_{\bar{t}_1}) = u.
\]
Since (\ref{L-lin-eq}) and (\ref{ttbar-lin-eq2}) are compatible,
the $L$-operators turn out to satisfy the Lax equations
\begin{gather}
  \frac{\rd L(\rd_{t_1})}{\rd\bar{t}_{2n+1}} =
  \bigl[\bar{Q}_{2n+1}(\rd_{t_1})^{-1}\bar{P}_{2n+1}(\rd_{t_1}),\,
        L(\rd_{t_1})\bigr],
  \nonumber \\
  \frac{\rd\bar{L}(\rd_{\bar{t}_1})}{\rd t_{2n+1}} =
  \bigl[Q_{2n+1}(\rd_{\bar{t}_1})^{-1}P_{2n+1}(\rd_{\bar{t}_1}),\,
        \bar{L}(\rd_{\bar{t}_1})\bigr].
  \label{Lax-eq2}
\end{gather}

\subsection{Quasi-classical approximation of wave functions}

We now consider the limit as $\hbar \to 0$. If the rescaled
$\tau$-function $\tau_\hbar(\bst,\bar{\bst})$ behaves like
(\ref{tau-F}), the associated wave functions
\[
  \Psi_\hbar(\bst,\bar{\bst},z) =
    \frac{\tau_\hbar(\bst-2\hbar[z^{-1}],\,\bar{\bst})}
         {\tau_\hbar(\bst,\bar{\bst})}
    e^{\hbar^{-1}\xi(\bst,z)}, \qquad
  \bar{\Psi}_\hbar(\bst,\bar{\bst},z) =
    \frac{\tau_\hbar(\bst,\bar{\bst}-2\hbar[z^{-1}])}
         {\tau_\hbar(\bst,\bar{\bst})}
    e^{\hbar^{-1}\xi(\bar{\bst},z)}
\]
turn out to take the quasi-classical form
\begin{gather}
  \Psi_\hbar(\bst,\bar{\bst},z) =
    \exp\bigl(\hbar^{-1}S(\bst,\bar{\bst},z) + O(1)\bigr),
  \qquad
  \bar{\Psi}_\hbar(\bst,\bar{\bst},z) =
    \exp\bigl(\hbar^{-1}\bar{S}(\bst,\bar{\bst},z) + O(1)\bigr).
  \label{Psi-S}
\end{gather}
To simplify notations, let us write the phase functions
$S(\bst,\bar{\bst},z)$, $\bar{S}(\bst,\bar{\bst},z)$ as
$S(z),\bar{S}(z)$. As expected, they have the Laurent expansion
\begin{gather*}
  S(z) =
    \sum_{n=0}^\infty t_{2n+1}z^{2n+1}
    - \sum_{n=0}^\infty
      \frac{2z^{-2n-1}}{2n+1}\rd_{t_{2n+1}}F,
  \\
  \bar{S}(z)=
    \sum_{n=0}^\infty \bar{t}_{2n+1}z^{2n+1}
    - \sum_{n=0}^\infty
      \frac{2z^{-2n-1}}{2n+1}\rd_{\bar{t}_{2n+1}}F
\end{gather*}
in a neighborhood of $z = \infty$. As the time variables are
rescaled, dif\/ferential operators in the auxiliary linear
equations are also rescaled as
\[
  \rd_{t_{2n+1}} \to \hbar\rd_{t_{2n+1}}, \qquad
  \rd_{\bar{t}_{2n+1}} \to \hbar\rd_{\bar{t}_{2n+1}}.
\]
Substituting the quasi-classical form (\ref{Psi-S}) of the wave
functions into those linear equations, we obtain a set of
Hamilton--Jacobi equations for the phase functions.

The fundamental auxiliary linear equations (\ref{t-lin-eq}),
(\ref{tbar-lin-eq}) and (\ref{ttbar-lin-eq}) thus yield the
Hamilton--Jacobi equations
\begin{gather}
  \rd_{t_{2n+1}}S(z) =
    \calB_{2n+1}\bigl(\rd_{t_1}S(z)\bigr), \nonumber\\
  \rd_{\bar{t}_{2n+1}}S(z) =
    \bar{\calB}_{2n+1}\bigl(\rd_{\bar{t}_1}S(z)\bigr), \nonumber\\
  \rd_{t_1}S(z)\cdot\rd_{\bar{t}_1}S(z) = u
  \label{ttbarS-u}
\end{gather}
for $S(z)$ and
\begin{gather*}
  \rd_{t_{2n+1}}\bar{S}(z) =
    \calB_{2n+1}\bigl(\rd_{t_1}\bar{S}(z)\bigr), \\
  \rd_{\bar{t}_{2n+1}}\bar{S}(z) =
    \bar{\calB}_{2n+1}\bigl(\rd_{\bar{t}_1}\bar{S}(z)\bigr), \\
  \rd_{t_1}\bar{S}(z)\cdot\rd_{\bar{t}_1}\bar{S}(z) = u
\end{gather*}
for $\bar{S}(z)$.   $\calB(p)$ and $\bar{\calB}(\bar{p})$ denote
the classical limit
\[
  \calB_{2n+1}(p) = \lim_{\hbar\to 0}B_{2n+1}(p),
  \qquad
  \bar{\calB}_{2n+1}(\bar{p})
   = \lim_{\hbar\to 0}\bar{B}_{2n+1}(\bar{p})
\]
of the ``symbols'' of the corresponding dif\/ferential operators.

Similarly, (\ref{L-lin-eq}) yields the equations
\[
  \calL(\rd_{t_1}S(z)) = z, \qquad
  \bar{\calL}(\rd_{\bar{t}_1}\bar{S}(z)) = z,
\]
where $\calL(p)$ and $\bar{\calL}(\bar{p})$, like
$\calB_{2n+1}(p)$ and $\bar{\calB}_{2n+1}(\bar{p})$, are given by
\[
  \calL(p) = \lim_{\hbar\to 0}L(p), \qquad
  \bar{\calL}(\bar{p}) = \lim_{\hbar\to 0}\bar{L}(\bar{p}).
\]
Thus $p(z) = \rd_{t_1}S(z)$ and $\bar{p}(z) =
\rd{\bar{t}_1}\bar{S}(z)$ turn out to be the inverse functions of
$z = \calL(p)$ and $z = \bar{\calL}(\bar{p})$, respectively.  The
relation (\ref{BBbar-LLbar}) connecting the $B$-operators and the
$L$-operators turns into the relation
\[
  \calB_{2n+1}(p) = \bigl(\calL(p)^{2n+1}\bigr)_{\ge 0},
  \qquad
  \bar{\calB}_{2n+1}(\bar{p})
   = \bigl(\bar{\calL}(\bar{p})^{2n+1}\bigr)_{\ge 0}
\]
among the corresponding phase space functions. $(\cdot)_{\ge 0}$
now stand for the polynomial part of Laurent series (of $p$ and
$\bar{p}$, respectively).  This, in particular, implies that
$\calB_{2n+1}(p)$ coincide with the Farber polynomials of the
inverse function $p = p(z)$ of $z = \calL(p)$.  A similar
statement holds for $\bar{\calB}_{2n+1}(\bar{p})$ and the inverse
function $\bar{p} = \bar{p}(z)$ of $z = \bar{\calL}(\bar{p})$.
Thus we have
\[
  \calB_{2n+1}(p) = \Phi_{2n+1}(p), \qquad
  \bar{\calB}_{2n+1}(\bar{p})
    = \bar{\Phi}_{2n+1}(\bar{p}).
\]

The somewhat strange linear equations (\ref{ttbar-lin-eq2}), too,
turn out to have a natural interpretation. The associated
Hamilton--Jacobi equations read
\begin{gather*}
  \bar{\calQ}_{2n+1}(\rd_{t_1}S(z))\rd_{\bar{t}_{2n+1}}S(z)
  = \bar{\calP}_{2n+1}(\rd_{t_1}S(z)),
  \nonumber \\
  \calQ_{2n+1}(\rd_{\bar{t}_1}S(z))\rd_{t_{2n+1}}S(z)
  = \calP_{2n+1}(\rd_{\bar{t}_1}S(z)),
\end{gather*}
for $S(z)$ and
\begin{gather*}
  \bar{\calQ}_{2n+1}(\rd_{t_1}\bar{S}(z))
  \rd_{\bar{t}_{2n+1}}\bar{S}(z)
  = \bar{\calP}_{2n+1}(\rd_{t_1}\bar{S}(z)),
  \nonumber \\
  \calQ_{2n+1}(\rd_{\bar{t}_1}\bar{S}(z))
  \rd_{t_{2n+1}}\bar{S}(z)
  = \calP_{2n+1}(\rd_{\bar{t}_1}S(z)),
\end{gather*}
$\bar{S}(z)$, where
\begin{gather*}
  \bar{\calQ}_{2n+1}(p)
    = \lim_{\hbar\to 0}\bar{Q}_{2n+1}(p), \qquad
  \bar{\calP}_{2n+1}(p)
    = \lim_{\hbar\to 0}\bar{P}_{2n+1}(p),
  \nonumber\\
  \calQ_{2n+1}(\bar{p})
    = \lim_{\hbar\to 0}Q_{2n+1}(\bar{p}), \qquad
  \calP_{2n+1}(\bar{p})
    = \lim_{\hbar\to 0}P_{2n+1}(\bar{p}).
\end{gather*}
For the equations of the $\bar{\bst}$-f\/lows to be consistent
with the previous Hamilton--Jacobi equations, $\calP_{2n+1}(p)$,
$\calQ_{2n+1}(p)$ and $\bar{\calB}_{2n+1}(\bar{p})$ have to
satisfy the equations
\begin{gather*}
  \bar{\calQ}_{2n+1}(\rd_{t_1}S(z))^{-1}
  \bar{\calP}_{2n+1}(\rd_{t_1}S(z))
  = \bar{\calB}_{2n+1}(\rd_{\bar{t}_1}S(z)),
  \nonumber \\
  \bar{\calQ}_{2n+1}(\rd_{t_1}\bar{S}(z))^{-1}
  \bar{\calP}_{2n+1}(\rd_{t_1}\bar{S}(z))
  = \bar{\calB}_{2n+1}(\rd_{\bar{t}_1}\bar{S}(z)).
\end{gather*}
Since $\rd_{t_1}S(z)$ and $\rd_{\bar{t}_1}S(z)$ obey the algebraic
relation (\ref{ttbarS-u}), these two equations reduce to a single
equation of the form
\begin{gather}
  \bar{\calB}_{2n+1}(u/p)
  = \bar{\calQ}_{2n+1}(p)^{-1}\bar{\calP}_{2n+1}(p).
  \label{Bbar-QbarPbar}
\end{gather}
Similarly, we have the relation
\begin{gather}
  \calB_{2n+1}(u/\bar{p})
  = \calQ_{2n+1}(\bar{p})^{-1}\calP_{2n+1}(\bar{p}).
  \label{B-QP}
\end{gather}
Actually, this is a result to be expected from
(\ref{ttbar-lin-eq2}), because (\ref{ttbar-lin-eq2}) can be
(formally) rewritten as
\begin{gather*}
  \bigl(\rd_{\bar{t}_{2n+1}}
      - \bar{Q}_{2n+1}(\rd_{t_1})^{-1}
        \bar{P}_{2n+1}(\rd_{t_1}) \bigr)\Psi = 0,
  \nonumber\\
  \bigl(\rd_{t_{2n+1}}
      - Q_{2n+1}(\rd_{\bar{t}_1})^{-1}
        P_{2n+1}(\rd_{\bar{t}_1}) \bigr)\Psi = 0,
\end{gather*}
and for these equations to be consistent with (\ref{t-lin-eq}) and
(\ref{tbar-lin-eq}), we should have
\begin{gather*}
  \bar{B}_{2n+1}(\rd_{\bar{t}_1})\Psi=
    \bar{Q}_{2n+1}(\rd_{t_1})^{-1}
    \bar{P}_{2n+1}(\rd_{t_1})\Psi, \\
  B_{2n+1}(\rd_{t_1})\Psi =
    Q_{2n+1}(\rd_{\bar{t}_1})^{-1}
    P_{2n+1}(\rd_{\bar{t}_1})\Psi,
\end{gather*}
which may be thought of as  ``quantization'' of
(\ref{Bbar-QbarPbar}) and (\ref{B-QP}).

\subsection{Dispersionless Lax equations}

In the limit as $\hbar \to 0$, (\ref{Lax-eq1}) and (\ref{Lax-eq2})
turn into the dispersionless Lax equations
\[
  \rd_{t_{2n+1}}\calL(p) =
  \{\calB_{2n+1}(p),\calL(p)\}_{p,t_1},
  \qquad
  \rd_{\bar{t}_{2n+1}}\bar{\calL}(\bar{p}) =
  \{\bar{\calB}_{2n+1}(\bar{p}),
    \bar{\calL}(\bar{p})\}_{\bar{p},\bar{t}_1}
\]
and
\begin{gather*}
  \rd_{\bar{t}_{2n+1}}\calL(p) =
  \{\bar{\calQ}_{2n+1}(p)^{-1}\bar{P}_{2n+1}(p),\,
    \calL(p)\}_{p,t_1},
  \nonumber\\
  \rd_{t_{2n+1}}\bar{\calL}(\bar{p}) =
  \{\calQ_{2n+1}(\bar{p})^{-1}\calP_{2n+1}(\bar{p}),\,
    \bar{\calL}(\bar{p})\}_{\bar{p},\bar{t}_1},
\end{gather*}
respectively.  In view of (\ref{Bbar-QbarPbar}) and (\ref{B-QP}),
we can rewrite the second two as
\[
  \rd_{\bar{t}_{2n+1}}\calL(p) =
  \{\bar{\calB}_{2n+1}(u/p),\calL(p)\}_{p,t_1},
  \qquad
  \rd_{t_{2n+1}}\bar{\calL}(\bar{p}) =
  \{\calB_{2n+1}(u/\bar{p}),
    \bar{\calL}(\bar{p})\}_{\bar{p},\bar{t}_1}.
\]
Obviously, these equations coincide with the corresponding
equations in (\ref{dLax-eq1}) and (\ref{dLax-eq2}).

\subsection*{Acknowledgements}

I would like to thank T.~Ikeda, T.~Shiota and T.~Takebe for
cooperation.  I also would like to thank L.~Martinez Alonso,
B.~Konopelchenko, M.~Ma\~{n}as and A.~Sorin for useful discussions
during the SISSA conference ``Riemann--Hilbert Problems,
Integrability and Asymptotics'' in September, 2005. This research
was partially supported by Grant-in-Aid for Scientif\/ic Research
No.~16340040 from the Japan Society for the Promotion of Science.

\LastPageEnding

\end{document}